\documentclass[nofootinbib]{revtex4-2}
\pdfoutput=1
\usepackage{graphicx}
\usepackage{dcolumn}
\usepackage{bm}
\usepackage{subfigure}
\usepackage{ulem}
\usepackage[table]{xcolor}
\usepackage{amsmath}
\usepackage{fancyvrb}
\usepackage[Lenny]{fncychap}
\usepackage{mathrsfs}
\usepackage{graphicx}  
\graphicspath{{img/}}
\usepackage{dcolumn}   
\usepackage{bm}        
\usepackage{amsmath,amssymb}
\usepackage{hyperref}
\hypersetup{colorlinks=true,linkcolor=red,filecolor=magenta,urlcolor=blue}
\usepackage{color}
\definecolor{purple}{rgb}{0.58,0.0,0.83}
\usepackage{caption}
\usepackage{tabularx}
\usepackage{comment}

\definecolor{owngreen}{rgb}{0.5, 0.5, 0.0}

\begin{document}


\title{Equivalence of Dark Energy Models: A Theoretical and Bayesian Perspective}

\author{David Tamayo$^{1,2}$}
\email{david.tr@piedrasnegras.tecnm.mx}

\author{Erick Urquilla$^{3,4}$}
\email{eurquill@vols.utk.edu}

\author{Isidro Gómez-Vargas$^{5}$}
\email{isidro.gomezvargas@unige.ch}

\affiliation{$^1$Instituto Tecnológico de Piedras Negras, 26080, Piedras Negras, Mexico}
\affiliation{$^2$Instituto de Astrof\'{\i}sica e Ci\^encias do Espa{\c c}o, Universidade do Porto, CAUP, 4150-762, Porto, Portugal}
\affiliation{$^3$ Department of Physics and Astronomy, University of Tennessee Knoxville, Knoxville, TN 37996, USA}
\affiliation{$^4$ Universidad de El Salvador, Final, 25 Av. Norte, San Salvador, El Salvador}
\affiliation{$^5$ Department of Astronomy of the University of Geneva, 51 Chemin Pegasi, 1290 Versoix, Switzerland.}


\begin{abstract}

We explore the background equivalence among three dark energy models by constructing explicit mappings between dynamical dark energy (DDE), interacting dark energy (IDE), and running vacuum (RV). In our approach, the dark sector functions that characterize each model—such as the equation of state parameter $\bar{w}(a)$ for DDE, the interaction term $Q$ for IDE, and the functional form $\Lambda(H)$ for RV—are transformed into one another under specific assumptions. 
Extending previous work by von Marttens et al. (2020), we demonstrate that running vacuum models, characterized by $\Lambda(H) = a_0 + a_1 \dot{H} + a_2 H^2$, can be reinterpreted as an interacting dark energy model with $Q = 3H\gamma \hat{\rho}_c$, which in turn is equivalent to a dynamic dark energy model with an appropriately defined $\bar{w}(a)$. Using Bayesian analysis with Type Ia supernovae, Baryon Acoustic Oscillations, and Cosmic Chronometers, our observational constraints confirm that these theoretical equivalences hold at the background level. 
This study underscores the importance of seeking convergence in dark energy models, facilitating a better understanding of the dark sector.
\end{abstract}

\maketitle

\section{Introduction}\label{sec: introduction}

Encompassing dark matter and dark energy, the dark sector is a cornerstone of the standard cosmological model. These components are believed to constitute approximately 95\% of the total energy density of the Universe, shaping its current dynamics and governing the formation of large-scale structures.  
However, their true nature remains one of the most profound mysteries in contemporary physics.
Dark matter is theorized to be a pressureless form of matter that plays a dominant role in the formation of cosmic structures.  
Its existence is inferred through its gravitational influence on visible matter, and extensive observational evidence supports its presence.  
Key indicators include discrepancies in galactic rotation curves, gravitational lensing phenomena, the large-scale distribution of galaxies, and the dynamics of colliding galaxy clusters.  
These observations collectively suggest the existence of significant amounts of unseen matter, estimated to comprise roughly 25\% of the Universe's total energy density \cite{Boddy:2022knd, Cirelli:2024ssz}.  
In contrast, dark energy is a hypothetical form of energy thought to permeate all of space and is regarded as the driver behind the observed accelerated expansion of the Universe.  
Accounting for approximately 70\% of the Universe's energy density, its influence is supported by multiple observations, including Type Ia supernovae, the distribution of large-scale structures, cosmic microwave background anisotropies, and baryon acoustic oscillations \cite{Huterer:2017buf}.  
There are numerous candidates for dark energy \cite{Bamba:2012cp, Motta:2021hvl, DiValentino:2021izs}; however, the simplest and most well-known is the cosmological constant term, denoted as $\Lambda$, which is introduced in the Einstein field equations. 
Together with cold dark matter, $\Lambda$ forms the foundation of the $\Lambda$CDM model, the prevailing paradigm in cosmology \cite{Planck:2018vyg}.  
Long-standing issues, such as the fine-tuning and coincidence problems associated with $\Lambda$, together with recent observational progress, have added uncertainty to the standard dark energy paradigm. DESI 2024 observations, combined with CMB data, prefer models with a time-varying dark energy equation of state, deviating by $2.6\sigma$ from $\Lambda$CDM. This tension persists or grows when the SN Ia data set is added \cite{DESI:2024mwx}. Dynamical dark energy models have shown moderate to strong Bayesian evidence preference over $\Lambda$CDM \cite{DESI:2024kob}. Evidence for dark energy evolution at low redshift suggests an emergent dark energy behaviour \cite{DESI:2024aqx,  Orchard:2024bve}. 
Other cosmological tensions, including discrepancies in the Hubble constant ($H_0$) and the matter clustering parameter ($S_8$) also offer significant challenges (see \cite{DiValentino:2021izs,Perivolaropoulos:2021jda, Abdalla:2022yfr, Vagnozzi:2023nrq} and references there).
These inconsistencies, observed between predictions from the $\Lambda$CDM model and independent measurements, suggest potential limitations of the standard framework and motivate the search for new physics better to understand the dark sector and the Universe's evolution \cite{Perivolaropoulos:2021jda, DiValentino:2021izs, Hu:2023jqc}.  
From a theoretical standpoint, one of the central challenges is understanding the fundamental nature of dark matter and dark energy.  
Efforts to reconcile the dark energy component with vacuum energy within the framework of quantum field theory have led to the cosmological constant problem, one of the most profound puzzles in modern physics \cite{Adler:2021arz, Lombriser:2019jia}.  
Similarly, the search for particles beyond the Standard Model to account for dark matter remains an active area of research \cite{Bertone:2018krk, Arbey:2021gdg}.  
 \\
 
Various models of the dark sector can produce identical cosmological background observables, a phenomenon known as \textit{dark degeneracy} \cite{Kunz:2007rk, Aviles:2011ak, Carneiro:2014uua, vonMarttens:2019ixw, Barbosa:2024ppn}.  
This concept arises because different theoretical descriptions of the dark sector—encompassing dark matter and dark energy—can lead to the same Hubble expansion rate, $H(z)$.  
The equivalence stems from the fact that the data through the Einstein's equations constrain only the total energy-momentum tensor, without offering separate constraints for the individual contributions of each component.  
To determine the specific role of each material component in cosmic expansion, additional information about their properties is required, such as their equation of state (EoS) and potential interactions with other components.  
For instance, the EoS for radiation ($w_r = 1/3$) and baryonic matter ($w_b = 0$) can be derived from statistical mechanics, and their energy densities evolve independently, without interaction. 
In the case of dark matter and dark energy, the $\Lambda$CDM model assumes EoS values of $w_c = 0$ and $w_x = -1$, respectively, with their energy densities also evolving independently.  
However, unlike radiation and baryonic matter, the fundamental nature of dark matter and dark energy is still unknown, and their EoS and potential interactions cannot yet be derived from first principles.  
As a result, the dark sector can be described by a wide variety of models, each consistent with current observational constraints.  
Due to dark degeneracy, these models can appear indistinguishable at the background level.  
For instance, in \cite{vonMarttens:2019ixw}, the authors demonstrate that this degeneracy makes it impossible to distinguish scenarios where dark energy interacts with dark matter from those involving non-interacting dynamical dark energy, using observational data based solely on time or distance measurements. 
However, the degeneracy is resolved at the perturbation level. 
Similarly, \cite{vonMarttens:2022xyr} shows that an interaction in the dark sector, modeled through the generalized Chaplygin gas, can effectively mimic a dynamical dark energy model.
This work extends the line of research by explicitly demonstrating the equivalence between dynamical dark energy, interacting dark energy, and running vacuum models. 
Building on the connection between dynamical and interacting dark energy established in \cite{vonMarttens:2019ixw}, we incorporate running vacuum into the framework. 
To validate the proposed equivalences, we perform a Bayesian analysis to constrain the parameters of each model, ensuring consistency with the theoretical predictions and providing a unified perspective on the dark sector.

In addition to the theoretical analysis, we conduct a Bayesian study to determine whether the equivalence among these models is also reflected in parameter estimation using background cosmological data. This allows us to obtain observational constraints for each model and perform a Bayesian model comparison, further assessing their consistency within the available data.

This paper is structured as follows:  
Section \ref{sec: theoretical background} establishes the theoretical cosmological framework utilized throughout the study. 
Section \ref{sec: dark energy models} introduces the three dark energy models under investigation: dynamical dark energy, interacting dark energy, and running vacuum.  
The results of this work are divided into two sections. 
Section \ref{sec: background equivalence between dark energy} demonstrates the theoretical equivalences between the selected models at the background level, while Section \ref{sec:bayesian} presents the observational constraints on the parameters of each model, illustrating consistency with the theoretical predictions.
Finally, Section \ref{sec: conclusions} discusses the results and presents the conclusions of this study.    

\section{Theoretical background}\label{sec: theoretical background}
In this work, we adopt the framework of a flat spacetime ($k = 0$) described by the Friedmann–Lemaître–Robertson–Walker (FLRW) line element within General Relativity ($c = 1$).  
The dynamics of the spacetime are governed by the Friedmann equations:  
\begin{eqnarray}
   \left(\frac{\dot{a}}{a}\right)^2 \equiv H^2 = \frac{8\pi G}{3} \rho, \label{friedmann 1}\\
    \frac{\ddot{a}}{a}= -\frac{4\pi G}{3} (\rho+3p), \label{friedmann 2}
\end{eqnarray}
where $\rho$ and $p$ represent the total energy density and pressure of the cosmic components, respectively, and $H = \frac{\dot{a}}{a}$ is the Hubble expansion rate.  
We assume four main components in the Universe: radiation, baryons, cold dark matter, and dark energy, denoted by the subscripts $r$, $b$, $c$, and $x$, respectively.  
Each component is modelled as a barotropic perfect fluid with EoS given by $p_i = w_i \rho_i$.  
Radiation and baryonic matter are characterized by $w_r = 1/3$ and $w_b = 0$, while cold dark matter is also pressureless, with $w_c = 0$.  
Despite sharing the same EoS, baryonic matter, and dark matter are treated as distinct components due to their different physical properties.  
Dark energy, on the other hand, is considered to have a general time-dependent EoS, $w_x(a)$.  

In the $\Lambda$CDM model, there is no interaction between the components of the Universe, meaning no exchange of energy or momentum occurs.  
We assume that baryons and radiation are independently conserved.  
For these components, the background energy conservation equation, $\dot{\rho}_i + 3H\rho_i = 0$, with EoS $w_b=0$ and $w_r = 1/3$ leads to:  
\begin{eqnarray}
    \rho_r &=& \rho_{r0} a^{-4}, \label{rho rad}\\
    \rho_b &=& \rho_{b0} a^{-3}. \label{rho bar}
\end{eqnarray}
For the dark sector, however, there is no compelling reason to assume independent conservation of energy densities.  
Thus, we allow for interactions between dark matter and dark energy.  
Following \cite{vonMarttens:2019ixw}, the dark sector can be described as a unified dark fluid ($d$) with energy density and pressure at the background level given by:  
\begin{eqnarray}
    \rho_d &=& \rho_c + \rho_x, \label{rho d}\\
    p_d &=& p_c + p_x = p_x = w_x(a) \rho_x, \label{p d}\\
    r &\equiv& \frac{\rho_c}{\rho_x} \label{r}.
\end{eqnarray}
Dark matter is assumed to be pressureless, and $r$ is the ratio of dark matter to dark energy densities.  
The unified dark fluid can be described by an effective EoS expressed in terms of the dark energy EoS parameter and the density ratio $r(a)$:  
\begin{eqnarray}
    p_d = \omega_d(a) \rho_d, \quad \omega_d(a) = \frac{\omega_x(a)}{1 + r(a)}.
\end{eqnarray}
The total energy density of the dark sector must be conserved.  
Thus, the background energy conservation equation can be written in terms of the effective EoS as:  
\begin{eqnarray}
    \dot{\rho}_d + 3H[1 + \omega_d(a)]\rho_d = 0. \label{conservation dark sector}
\end{eqnarray}
Until a specific model is specified to separate the components of the dark sector, its dynamics at the background level are governed by the effective EoS parameter $\omega_d(a)$.  
The phenomenon of dark degeneracy arises from the fact that (see \eqref{conservation dark sector}) $\omega_d(a)$ depends on both the dark energy EoS parameter $\omega_x(a)$, which reflects the potential dynamic nature of dark energy independent of other components, and $r(a)$, which accounts for possible interactions between dark matter and dark energy.  
As a result, different combinations of $\omega_x(a)$ and $r(a)$ can produce the same $\omega_d(a)$, and consequently, the same Hubble expansion rate $H$. 
It is important to emphasize that dark degeneracy does not imply that two descriptions with the same $\omega_d(a)$ but different $\omega_x(a)$ and $r(a)$ are identical.  
The difference lies in how dark matter and dark energy evolve: while the two descriptions lead to the same $H$, their components follow distinct evolutionary paths.  
This has a significant implication: observations based solely on distance measurements—such as those using Type Ia supernovae, Baryon Acoustic Oscillations, or cosmic chronometers—cannot distinguish between models with identical $\omega_d(a)$.  
However, at the perturbation level, dark degeneracy can be broken. 

\section{Dark energy models}\label{sec: dark energy models}

This study focuses on three well-known dark energy models within the context of General Relativity: dynamical dark energy (DDE), interacting dark energy (IDE) and running vacuum (RV).  

We will briefly summarize them.  

\begin{itemize}
\item \textbf{Dynamical dark energy} - Unlike $\Lambda$CDM, where the dark energy EoS is constant ($w_x = -1$), in DDE models the EoS is dynamic, meaning it depends on the evolution of the Universe, $\bar{w}_x(a)$ (quantities associated with DDE are labelled with a bar).  
Many DDE parameterizations have been extensively studied \cite{Copeland:2006wr, Dai:2018zwv, Vazquez:2020ani, Colgain:2021pmf, Yang:2022klj, DESI:2024mwx}.  

In DDE, each cosmic component evolves independently according to the conservation equation:  
\begin{equation}\label{dde conservation}
    \dot{\bar{\rho}}_{i} +3H\left(1 +\bar{w}_{i} \right)\bar{\rho}_{i}=0,
\end{equation}
where the EoS parameters are $\{\bar{w}_{r} = 1/3, \, \bar{w}_{b} = \bar{w}_{c} = 0,\, \bar{w}_x = \bar{w}_x(a)\}$.  
Integrating \eqref{dde conservation} for dark matter and dark energy yields:  
\begin{eqnarray}
    \bar{\rho}_c &=& \bar{\rho}_{c0}\, a^{-3}, \label{dde rhoc}\\
    \bar{\rho}_x &=& \bar{\rho}_{x0}\, \exp \left({-3\int \frac{1+\bar{w}_x(a)}{a}da}\right). \label{dde rhox}
\end{eqnarray}
The dark matter energy density evolves in the same way as baryonic matter in $\Lambda$CDM.  
However, the dark energy density, as seen from \eqref{dde rhox}, strongly depends on the specific form of the dark energy EoS $\bar{w}_x(a)$.  
The key difference between DDE and $\Lambda$CDM lies in the description of dark energy.  

\item \textbf{Interacting dark energy} - The IDE models retain the same EoS parameters as $\Lambda$CDM (IDE quantities will be denoted with a tilde): $\{\tilde{w}_{r} = 1/3, \, \tilde{w}_{b} = \tilde{w}_{c} = 0,\, \tilde{w}_x = -1\}$.  
While the radiation and baryonic matter components are assumed to evolve independently and conserve their energy, the dark matter and dark energy components are not independently conserved.  
The central hypothesis of IDE models is that dark matter and dark energy are not isolated entities but interact by exchanging energy and momentum with one another \cite{Costa:2016tpb, vonMarttens:2018iav, Yang:2018euj, DiValentino:2019ffd, DiValentino:2019jae,Aljaf:2020eqh, Pan:2020mst, Guo:2021rrz, Kaeonikhom:2022ahf, Mishra:2023ueo, Hoerning:2023hks, Giare:2024smz}.  
For a comprehensive review, see \cite{Bolotin:2013jpa, Wang:2016lxa, Wang:2024vmw}.  

This interaction is captured through coupled energy conservation equations:  
\begin{eqnarray}
    \dot{\tilde{\rho}}_{c} + 3H\tilde{\rho}_c &=& \Tilde{Q}, \label{ide cons1}\\
    \dot{\tilde{\rho}}_x &=& -\Tilde{Q},  \label{ide cons2}
\end{eqnarray}
where $\Tilde{Q}$ is a scalar function that phenomenologically quantifies the energy exchange between dark matter and dark energy.  
The distinction between IDE models and $\Lambda$CDM lies in the inclusion of this interaction term, $\Tilde{Q}$, which must be specified to determine the evolution of the energy densities.  

\item \textbf{Running vacuum} - The RV models are motivated by the renormalization group in quantum field theory, which suggests treating the vacuum as a dynamic entity.  
In this framework, $\Lambda$ is not a constant but evolves over time as $\Lambda(H)$ \cite{Shapiro:2009dh, Gomez-Valent:2014rxa, Rezaei:2021qwd, Sola:2016zeg, Sola:2016jky,SolaPeracaula:2021gxi, SolaPeracaula:2022hpd, SolaPeracaula:2023swx}.  
The general form of $\Lambda$ can be expressed as:  
\begin{equation}\label{rv general}
\Lambda = a_{0} + \sum_{k} b_{k} H^{2k} + \sum_{k} c_{k} \dot{H}^{k}.
\end{equation}
In RV models, the EoS parameters are the same as those in $\Lambda$CDM (quantities in RV will be denoted with a hat): $\{\hat{w}_{r} = 1/3, \, \hat{w}_{b} = \hat{w}_{c} = 0,\, \hat{w}_x = -1\}$.  
Here, dark energy arises from the dynamics of $\Lambda$, and since $\Lambda$ depends on $H(a)$, it may seem natural to interpret RV as a subset of DDE.  
While this interpretation holds to some extent, as will be demonstrated later, certain specific cases of RV models establish a direct connection with the IDE framework.  
\end{itemize}
\section{Background equivalence between dark energy models}\label{sec: background equivalence between dark energy}

\subsection{Equivalence between DDE and IDE}\label{subsec: equivalence dde ide}

A detailed description of the dark sector degeneracy in the context of DDE and IDE models and how the degeneracy is broken at the linear level is provided in von Marttens et al. \cite{vonMarttens:2019ixw}.  
Building on the work of these authors, for specific forms of $\tilde{Q}$, it is possible to rewrite the system of equations \eqref{ide cons1} and \eqref{ide cons2} in terms of the ratio $\tilde{r} = \tilde{\rho}_c / \tilde{\rho}_x$.  
The transformed equations become: 
\begin{eqnarray}
    \dot{\tilde{\rho}}_c + 3H\tilde{\rho}_c \left(1 - \frac{\tilde{f}(\tilde{r})}{1 + \tilde{r}}\right) &=& 0,\\
    \dot{\tilde{\rho}}_x + 3H\tilde{\rho}_x \left(\frac{\tilde{f}(\tilde{r})}{1 + \tilde{r}^{-1}}\right) &=& 0,\\
    \text{where} \quad \tilde{f}(\tilde{r}) &=& \frac{\tilde{Q}}{3H}\left(\frac{\tilde{\rho}_c + \tilde{\rho}_x}{\tilde{\rho}_c \tilde{\rho}_x}\right). \label{f Q}
\end{eqnarray}
Taking the derivative of $\tilde{r} = \tilde{\rho}_c / \tilde{\rho}_x$ and performing some algebra, one finds that $\tilde{f}(\tilde{r})$ satisfies:
\begin{eqnarray}
    \tilde{f}(\tilde{r}) = \frac{a \tilde{r}'}{3 \tilde{r}} + 1, \label{f}
\end{eqnarray}
where the prime denotes differentiation with respect to the scale factor.  
This function $\tilde{f}(\tilde{r})$ is crucial for relating IDE models to DDE models.  

Given a DDE EoS parameter $\bar{\omega}_x(a)$, the energy densities of the dark components can be determined.  
The total unified dark sector energy density must be consistent between the two approaches, leading to:
\begin{equation}
    \rho_d = \bar{\rho}_c + \bar{\rho}_x = \tilde{\rho}_c + \tilde{\rho}_x.
\end{equation}
For both DDE and IDE, dark matter is considered pressureless ($\omega_c = 0$).
Using \eqref{rho d}, \eqref{p d} and \eqref{r} and rearranging terms, the EoS for the dark sector is given by:  
\begin{equation}
    \omega_d(a) = \frac{\omega_x(a) \rho_x}{\rho_c + \rho_x}.
\end{equation}
For IDE, the dark energy EoS is fixed ($\omega_x(a) \to \tilde{\omega}_x = -1$), yielding  
\begin{equation}
    \tilde{\omega}_d(a) = \frac{-\tilde{\rho}_x}{\tilde{\rho}_c + \tilde{\rho}_x}.
\end{equation}
For DDE, the dark energy EoS is dynamical ($\omega_x(a) \to \bar{\omega}_x(a)$), resulting in: 
\begin{equation}
    \bar{\omega}_d(a) = \frac{\bar{\omega}_x(a) \bar{\rho}_x}{\bar{\rho}_c + \bar{\rho}_x}.
\end{equation}

A crucial insight from the dark degeneracy is that the information about the dark sector is encoded in $\omega_d(a)$, which is degenerate in terms of combinations of dark energy and dark matter.  
Thus, the last two expressions must satisfy $\bar{\omega}_d(a) = \tilde{\omega}_d(a)$, leading to the following relations:  
\begin{eqnarray}
    \tilde{\rho}_x &=& -\bar{\omega}_x(a) \bar{\rho}_x, \label{rhox dde ide}\\
    \tilde{\rho}_c &=& \bar{\rho}_c + \left(1 + \bar{\omega}_x(a)\right) \bar{\rho}_x. \label{rhoc dde ide}
\end{eqnarray}
These equations reflect the equivalence between IDE and DDE models at the background level.  
Using \eqref{rhoc dde ide} and \eqref{rhox dde ide}, they can be solved for the background equations of the equivalent model in either approach.  
Once this equivalence is established, both the Hubble expansion rate and the unified dark fluid EoS will have identical forms in the two models.  
It is important to note, however, that the present values of the dark matter and dark energy density parameters differ: $\bar{\rho}_{c0} \neq \tilde{\rho}_{c0}$ and $\bar{\rho}_{x0} \neq \tilde{\rho}_{x0}$.  
Finally, combining \eqref{rhoc dde ide} and \eqref{rhox dde ide}, the DDE EoS can be related to IDE quantities as follows:  
\begin{eqnarray}
    \bar{\omega}_x = -1 + \frac{\tilde{\rho}_c - \bar{\rho}_c}{\tilde{\rho}_x + \tilde{\rho}_c - \bar{\rho}_c}. \label{dde ide w}
\end{eqnarray}
This equation establishes the connection between the DDE EoS and IDE quantities, further demonstrating their equivalence under specific conditions.  

To illustrate how this formalism works, let us consider a simple example.  
We take the simplest DDE model, the flat $w$CDM model, where $\bar{\omega}(a) = \omega_0$ is constant.  
In this case, the energy densities of the dark sector are given by:  
\begin{eqnarray}
    \bar{\rho}_c = \bar{\rho}_{c0} a^{-3},\quad \bar{\rho}_x = \bar{\rho}_{x0} a^{-3(1+\omega_0)}.
\end{eqnarray}
Using \eqref{rhoc dde ide} and \eqref{rhox dde ide} in the equations above, we can easily compute the corresponding IDE energy densities and the dark sector ratio:  
\begin{eqnarray}
    \tilde{\rho}_c &=& \bar{\rho}_{c0} a^{-3} + (1+\omega_0)\bar{\rho}_{x0} a^{-3(1+\omega_0)},\\
    \tilde{\rho}_x &=& -\omega_0 \bar{\rho}_{x0} a^{-3(1+\omega_0)},\\
    \tilde{r} &=& -\frac{1+\omega_0}{\omega_0} - \frac{\tilde{r}_0}{\omega_0} a^{3\omega_0}.
\end{eqnarray}
It is important to note that the energy densities of the dark components today differ between the two models, but their total energy density remains the same. Specifically:  
$\tilde{\rho}_{c0} = \bar{\rho}_{c0} + (1+\omega_0)\bar{\rho}_{x0}$ and $\tilde{\rho}_{x0} = -\omega_0 \bar{\rho}_{x0}$.  
With the expression for $\tilde{r}$, we can calculate $\tilde{r}'$ and use \eqref{f} to determine $\tilde{f}$. After some algebra, we find:  
\begin{eqnarray}
    \tilde{f} = (1+\omega_0)\left(1+\frac{1}{\tilde{r}}\right) = (1+\omega_0)\left(\frac{\tilde{\rho}_c + \tilde{\rho}_x}{\tilde{\rho}_c}\right).
\end{eqnarray}
Finally, using \eqref{f Q}, we can calculate the IDE interaction term $\tilde{Q}$:  
\begin{eqnarray}
    \tilde{Q} = 3H(1+\omega_0)\tilde{\rho}_x.
\end{eqnarray}
Thus, we have demonstrated that the flat FLRW model with pressureless dark matter and DDE characterized by $\bar{\omega} = \omega_0$ is background-equivalent to a flat FLRW model with IDE ($\omega_x = -1$) with interaction term is $\tilde{Q} = 3H\gamma \tilde{\rho}_x= 3H(1+\omega_0)\tilde{\rho}_x$.  

\subsection{Equivalence between RV and IDE}\label{subsec: equivalence rv ide}

The general form of $\Lambda$ in RV models is given in \eqref{rv general}.  
In this work, we focus on the first two non-trivial cases, denoted as A1 and A2, following the notation in \cite{Gomez-Valent:2014rxa}:  
\begin{eqnarray}
A1: \hspace{0.5cm} \Lambda &=& a_{0} + a_{2}H^{2}, \label{A1}\\
A2: \hspace{0.5cm} \Lambda &=& a_{0} + a_{1}\dot{H} + a_{2}H^{2}. \label{A2}
\end{eqnarray}
The A1 and A2 models have been extensively studied, demonstrating their viability as dark energy candidates, supported by observations \cite{Sola:2016jky, Sola:2016zeg}, and their potential to alleviate cosmological tensions \cite{SolaPeracaula:2023swx}.  
Considering these two RV models, we analyse a late-time Universe described by a flat FLRW spacetime filled with pressureless dark matter ($\hat{\rho}_c$) and dark energy ($\hat{\rho}_x = \Lambda / 8\pi G$), where the dark energy EoS satisfies $\hat{\rho}_x = -\hat{p}_x$.  
The Friedmann equations for this scenario are:  
\begin{eqnarray}
    H^{2} &=& \frac{8\pi G}{3}\left(\hat{\rho}_c + \hat{\rho}_x\right), \label{rv friedmann1}\\
    \dot{H} &=& -4\pi G \hat{\rho}_c. \label{rv friedmann2}
\end{eqnarray}
Substituting \eqref{rv friedmann1} and \eqref{rv friedmann2} into \eqref{A1} and \eqref{A2}, we derive the dark energy densities for each model:  
\begin{eqnarray}
A1: \hspace{0.5cm} \hat{\rho}_x &=& \left(\frac{3}{3-a_{2}}\right)\left(\frac{a_{0}}{8\pi G} + \frac{a_{2}}{3}\hat{\rho}_c\right),\\
A2: \hspace{0.5cm} \hat{\rho}_x &=& \left(\frac{3}{3-a_{2}}\right)\left(\frac{a_{0}}{8\pi G} + \left(\frac{a_{2}}{3} - \frac{a_{1}}{2}\right)\hat{\rho}_c\right).
\end{eqnarray}
For both models, the dark energy density can be expressed in the general form:  
\begin{equation}\label{rv de}
    \hat{\rho}_x = \alpha_{0} + \alpha_{1}\hat{\rho}_c,
\end{equation}
where the parameters $\alpha_{0}$ and $\alpha_{1}$ are given by:  
\begin{eqnarray}
A1: \hspace{0.5cm} \alpha_{0} &=& \frac{1}{8\pi G}\left(\frac{3a_{0}}{3-a_{2}}\right), \hspace{0.5cm}
\alpha_{1} = \frac{a_{2}}{3}\left(\frac{3}{3-a_{2}}\right), \label{A1 alpha}\\
A2: \hspace{0.5cm} \alpha_{0} &=& \frac{1}{8\pi G}\left(\frac{3a_{0}}{3-a_{2}}\right), \hspace{0.5cm}
\alpha_{1} = \left(\frac{a_{2}}{3} - \frac{a_{1}}{2}\right)\left(\frac{3}{3-a_{2}}\right). \label{A2 alpha}
\end{eqnarray}
It is important to emphasize that since we are considering a universe composed solely of dark matter and dark energy (as described by equation \eqref{rv friedmann1}), the specific RV models presented in equations \eqref{A1} and \eqref{A2} allow for the linear relationship between the components of the dark sector given by equation \eqref{rv de}. 
This analysis only applies to the late universe, where the contribution of radiation becomes negligible.

To derive the explicit expressions for dark matter and dark energy densities, we start from the general conservation equation $\sum_i [\dot{\hat{\rho}}_i + 3H(1+\hat{\omega}_i)\hat{\rho}_i] = 0$. For the dark sector, this equation simplifies to:   
\begin{equation}
\dot{\hat{\rho}}_c + 3H\hat{\rho}_c + \dot{\hat{\rho}}_x = 0. \label{rv conservation de dm}
\end{equation}
Substituting \eqref{rv de} into \eqref{rv conservation de dm} and rearranging, we find:  
\begin{eqnarray}
(1+\alpha_1)\dot{\hat{\rho}}_c + 3H\hat{\rho}_c &=& 0, \\
\Rightarrow \quad \frac{d\hat{\rho}_c}{da} &=& -\left(\frac{3}{1+\alpha_1}\right)\frac{\hat{\rho}_c}{a}.
\end{eqnarray}
Integrating, the dark matter density is obtained as:  
\begin{equation}
    \hat{\rho}_c = \hat{\rho}_{c0} a^{-3/(1+\alpha_{1})}. \label{rv rhoc}
\end{equation}

To calculate the dark energy density, we substitute \eqref{rv rhoc} in \eqref{rv de}. 
Using $\hat{\rho}_x(a=1) \equiv \hat{\rho}_{x0}$, we can eliminate one constant, $\alpha_{0} = \hat{\rho}_{x0} - \alpha_{1}\hat{\rho}_{c0}$, yielding the dark energy density:
\begin{equation}
    \hat{\rho}_x = \hat{\rho}_{x0} + \alpha_{1}\hat{\rho}_{c0}\left(a^{-3/(1+\alpha_{1})} - 1\right). \label{rv rhox}
\end{equation}
It is clear that for $\alpha_{1} = 0$, equations \eqref{rv rhoc} and \eqref{rv rhox} reduce to the $\Lambda$CDM solutions.  
From \eqref{rv rhoc} and \eqref{rv rhox}, we calculate $E \equiv \frac{H^{2}}{H_0^2} = \frac{8\pi G}{3 H_0^2}(\hat{\rho}_{c} + \hat{\rho}_{x})$:
\begin{eqnarray}
    \hat{E} = 1 + (1+\alpha_{1})\hat{\Omega}_{c0}\left(a^{-3/(1+\alpha_{1})} - 1\right). \label{rv E}
\end{eqnarray}

To show the equivalence between A1 and A2 RV models and IDE is straightforward.  
First, notice that the conservation equation \eqref{rv conservation de dm}, $\dot{\hat{\rho}}_c + 3H\hat{\rho}_c + \dot{\hat{\rho}}_x = 0$, can be split into two equations:
\begin{eqnarray}
    \dot{\hat{\rho}}_{c} + 3H\hat{\rho}_c &=& \hat{Q}, \label{rv ide1}\\
    \dot{\hat{\rho}}_x &=& -\hat{Q}. \label{rv ide2}
\end{eqnarray}
This gives the form of IDE models, indicating energy transfer between the vacuum (dark energy) and dark matter, showing that dark energy decays into matter or radiation, depending on the model, throughout the evolution of the Universe.
Computing $\dot{\hat{\rho}}_x$ from \eqref{rv de} and \eqref{rv rhoc}, we find the function $Q$:
\begin{eqnarray}
    \dot{\hat{\rho}}_x = \alpha_1 \dot{\hat{\rho}}_c = -3H\left(\frac{\alpha_{1}}{1+\alpha_{1}}\right)\hat{\rho}_c,\\
    \Rightarrow \quad \hat{Q} = 3H\left(\frac{\alpha_{1}}{1+\alpha_{1}}\right)\hat{\rho}_c. \label{rv Q}
\end{eqnarray}

We will make some observations about this result.  
Under the assumptions of the A1 and A2 RV models, the dark energy density is linearly proportional to the dark matter content, $\hat{\rho}_x = \alpha_{0} + \alpha_{1} \hat{\rho}_c$ (eq. \eqref{rv de}).  
This indicates interaction between the dark components.  
This interaction becomes explicit when the conservation equation is split, yielding eq. \eqref{rv Q}, which shows that the IDE coupling is proportional to the Hubble parameter and the dark matter density, $Q \propto H \hat{\rho}_c$.  
This is a well-known IDE model studied by several authors (for recent examples, see \cite{Guo:2021rrz, Kaeonikhom:2022ahf, Mishra:2023ueo}).  
The fact that RV models can be written as IDE has been noted previously by \cite{Perico:2016kbu, De-Santiago:2016oeu, Sola:2016zeg}, which is why RV models are sometimes called ``vacuum decay" models.  
This is due to the characteristic that $\Lambda$ is dynamical and transfers (or acquires, depending on the case) energy from another cosmic component.  
Here we explicitly show how two particular RV models are equivalent to an IDE model.

For completeness, following we demonstrate the equivalence in the reverse direction.  
Starting with an IDE model with $\tilde{\rho}_{\text{tot}} = \tilde{\rho}_c + \tilde{\rho}_x$ and EoS parameters $\tilde{w}_c = 0$ and $\tilde{w}_x = -1$, the conservation equations are:
\begin{eqnarray}
    \dot{\tilde{\rho}}_c + 3H\tilde{\rho}_c &=& \tilde{Q}, \label{IDE cons c}\\
    \dot{\tilde{\rho}}_x &=& -\tilde{Q}. \label{IDE cons x}
\end{eqnarray}
Now, let us assume the specific IDE model:
\begin{equation}
    \tilde{Q} = 3H\gamma \tilde{\rho}_c. \label{IDE Q}
\end{equation}
Combining \eqref{IDE cons c} and \eqref{IDE Q}, we find:
\begin{eqnarray}
    \dot{\tilde{\rho}}_c + 3H\tilde{\rho}_c &=& 3H\gamma \tilde{\rho}_c,\\
    &\Rightarrow& \quad \tilde{\rho}_c = \tilde{\rho}_{c0}a^{-3(1-\gamma)}. \label{IDE rhoc}
\end{eqnarray}
Now we solve \eqref{IDE cons x}:
\begin{eqnarray}
    \dot{\tilde{\rho}}_x &=& -3H\gamma \tilde{\rho}_c = -3H \tilde{\rho}_{c0}a^{-3(1-\gamma)},\\
    &\Rightarrow& \quad \tilde{\rho}_x = \tilde{\rho}_{x0} + \frac{\gamma}{1-\gamma}\tilde{\rho}_{c0}\left(a^{-3(1-\gamma)} - 1\right). \label{IDE rhox}
\end{eqnarray}
Finally, the normalized Hubble parameter becomes:
\begin{eqnarray}
    \tilde{E} = 1 + \frac{\tilde{\Omega}_{c0}}{1-\gamma}\left(a^{-3(1-\gamma)} - 1\right). \label{IDE E}
\end{eqnarray}
It is easy to see that if we set $\gamma = \frac{\alpha_1}{1+\alpha_1}$, then $\tilde{Q} = \hat{Q}$ and consequently $\tilde{\rho}_c = \hat{\rho}_c$ and $\tilde{\rho}_x = \hat{\rho}_x$, proving that both models are mathematically equivalent at the background level.

\subsection{Equivalence between RV and DDE}\label{subsec: equivalence rv dde}

Considering established that the A1 and A2 RV models provide a relationship between the dark sector components, $\hat{\rho}_x = \alpha_0 + \alpha_1 \hat{\rho}_c$, and their equivalence to an IDE model with the interaction term $\hat{Q} = 3H\frac{\alpha_1}{1+\alpha_1} \hat{\rho}_c$, we can use the DDE-IDE relations presented in section \ref{subsec: equivalence dde ide} to determine the corresponding DDE form.  
We start with a DDE model where $\bar{\rho}_{\text{tot}} = \bar{\rho}_c + \bar{\rho}_x$ and the EoS parameters are $\bar{w}_c = 0$ and $\bar{w}_x = \bar{w}(a)$.  
If there is no interaction between the dark sector components, the conservation equation for dark matter is $\dot{\bar{\rho}}_c + 3H\bar{\rho}_c = 0$, which integrates straightforwardly to give the standard form $\bar{\rho}_c = \bar{\rho}_{c0}a^{-3}$.  

To proceed, we must find the explicit form of the EoS parameter $\bar{w}(a)$.  
For this, we use the relationship between the EoS parameter and energy densities shown in \eqref{dde ide w}, substituting the energy densities $\tilde{\rho}_c$ from \eqref{IDE rhoc} and $\tilde{\rho}_x$ from \eqref{IDE rhox} (alternatively, the same result is obtained using $\hat{\rho}_c$ and $\hat{\rho}_x$).  
After some algebra, we obtain an explicit form for the EoS parameter:
\begin{eqnarray}
    \bar{\omega}(a) = -1 + \frac{\tilde{\Omega}_{c0}a^{-3(1-\gamma)} - \bar{\Omega}_{c0}a^{-3}}{\left[1 + \frac{\tilde{\Omega}_{c0}}{1-\gamma}\left(a^{-3(1-\gamma)} - 1\right) - \bar{\Omega}_{c0}a^{-3}\right]}. \label{DDE w}
\end{eqnarray}

Clearly, for $\gamma = 0$, i.e., when there is no interaction in the dark sector or equivalently no RV, the energy densities reduce to $\hat{\rho}_c = \hat{\rho}_{c0}a^{-3}$ and $\hat{\rho}_x = \hat{\rho}_{x0}$, implying $\hat{\Omega}_{c0} = \bar{\Omega}_{c0}$ and $\hat{\Omega}_{x0} = \bar{\Omega}_{x0}$. 
Consequently, the dark energy EoS simplifies to $\bar{\omega}(a) = -1$.  
Examining the form of the EoS parameter $\bar{\omega}(a)$ in \eqref{DDE w}, the most notable feature is the potential presence of a pole. This phenomenon is not entirely unexpected, as divergences in the dark energy EoS have been previously reported in other plausible dark energy models \cite{Gavela:2009cy, Akarsu:2019ygx, Akarsu:2019hmw}. 

With the explicit form of $\bar{w}(a)$, we can integrate the conservation equation for dark energy, $\dot{\bar{\rho}}_x + 3H(1+\bar{w}(a))\bar{\rho}_x = 0$, to obtain:
\begin{eqnarray}
    \bar{\rho}_x = \frac{\bar{\rho}_{x0}}{1-\bar{\Omega}_{c0}}\left[1 - \bar{\Omega}_{c0}a^{-3} + \frac{\tilde{\Omega}_{c0}}{1-\gamma}\left(a^{-3(1-\gamma)} - 1\right)\right]. \label{DDE rhox}
\end{eqnarray}
Then, we easily calculate $\bar{E} \equiv \frac{8\pi G}{3 H_0^2}(\bar{\rho}_c + \bar{\rho}_x)$. 
Note that $\bar{E} = \tilde{E}$ (see \eqref{IDE E}).
\begin{eqnarray}
    \bar{E}^2 = 1 + \frac{\tilde{\Omega}_{c0}}{1-\gamma}\left(a^{-3(1-\gamma)} - 1\right).
\end{eqnarray}

\subsection{Equivalence between dark energy models}\label{subsec: equivalence between dark energy models}

In Section \ref{subsec: equivalence dde ide}, we outline the formalism developed by von Marttens et al. \cite{vonMarttens:2019ixw}, which demonstrates how the functions characterizing DDE, $\{\bar{\rho}_x, \bar{\rho}_c, \bar{w}(a)\}$, can be transformed into the corresponding IDE functions $\{\tilde{\rho}_x, \tilde{\rho}_c, \tilde{Q}\}$. 
The transformation DDE-IDE is given by the equations \eqref{rhox dde ide}, \eqref{rhoc dde ide}, and \eqref{dde ide w}. 

In Section \ref{subsec: equivalence rv ide}, we extend this framework to the A1 \eqref{A1} and A2 \eqref{A2} RV models. 
In the late universe, where the dark sector dominates, these models exhibit a linear relationship between their components, $\hat{\rho}_x = \alpha_0 + \alpha_1 \hat{\rho}_c$ \eqref{rv de}. 
This relationship enables the derivation of the IDE-RV equivalence through the interacting term $\hat{Q} = 3H\left(\frac{\alpha_1}{1+\alpha_1}\right)\hat{\rho}_c$, shown in equation \eqref{rv Q}. 
Consequently, the RV functions $\{\hat{\rho}_x, \hat{\rho}_c, \Lambda(H)\}$ can be transformed into the IDE functions $\{\tilde{\rho}_x, \tilde{\rho}_c, \hat{Q}(\alpha_1)\}$.
Finally, in Section \ref{subsec: equivalence rv dde}, we complete the equivalence by deriving the RV-DDE relation. 
Since A1 and A2 RV models can be written as IDE models in the late universe, the equivalent DDE is determined by calculating the dynamical EoS parameter $\bar{w}(a)$ using equation \eqref{dde ide w}, yielding the expression \eqref{DDE w}. 
This establishes the full equivalence between the three models, linking their defining functions and parameters.

\begin{table}[h!]
    \centering
    \resizebox{\textwidth}{!}{
    \begin{tabular}{|c|c|c|c|}
        \hline
         & DDE & IDE & RV \\
         \hline \hline
        \textbf{Model} & 
        $\bar{\omega}(a) = -1 + \frac{\tilde{\Omega}_{c0}a^{-3(1-\gamma)} - \bar{\Omega}_{c0}a^{-3}}{\left[1 + \frac{\tilde{\Omega}_{c0}}{1-\gamma}\left(a^{-3(1-\gamma)} - 1\right) - \bar{\Omega}_{c0}a^{-3}\right]}$ & 
        $Q = 3H\gamma \hat{\rho}_c$ & 
        $\Lambda(H) = a_0 + a_1\dot{H} + a_2H^2$ \\ 
        \hline 
        \textbf{DM} & 
        $\bar{\rho}_c = \bar{\rho}_{c0} a^{-3}$ & 
        $\tilde{\rho}_{c} = \tilde{\rho}_{c0} a^{-3(1-\gamma)}$ & 
        $\hat{\rho}_{c} = \hat{\rho}_{c0} a^{-3/(1+\alpha_{1})}$ \\ 
        \hline 
        \textbf{DE} & 
        $\bar{\rho}_x = \frac{\bar{\rho}_{x0}}{1-\bar{\Omega}_{c0}}\left[1 - \bar{\Omega}_{c0}a^{-3} + \frac{\tilde{\Omega}_{c0}}{1-\gamma}\left(a^{-3(1-\gamma)} - 1\right)\right]$ & 
        $\tilde{\rho}_x = \tilde{\rho}_{x0} + \frac{\gamma}{1-\gamma}\tilde{\rho}_{c0}\left(a^{-3(1-\gamma)} - 1\right)$ & 
        $\hat{\rho}_x = \hat{\rho}_{x0} + \alpha_{1}\hat{\rho}_{c0}\left(a^{-3/(1+\alpha_{1})} - 1\right)$ \\ 
        \hline
        \textbf{EoS} & 
        $\bar{w}_c = 0$, $\bar{w}_x = \bar{w}(a)$ & 
        $\tilde{w}_c = 0$, $\tilde{w}_x = -1$ & 
        $\hat{w}_c = 0$, $\hat{w}_x = -1$ \\ 
        \hline
    \end{tabular}
    }
    \caption{Summary of the equivalences between the DDE, IDE, and RV models. The first row presents the defining equations for each model: the EoS parameter $\bar{w}(a)$ for DDE, the interaction term $Q$ for IDE, and $\Lambda(H)$ for RV. The subsequent rows include the explicit expressions for the dark matter $\rho_c$ and dark energy $\rho_x$ densities, as well as the corresponding EoS parameters for each framework.}
    \label{tab:model_equivalence}
\end{table}

Table \ref{tab:model_equivalence} summarizes the results obtained in this section and highlights the equivalences between the dark sector components in the DDE, IDE, and RV models. 
For each of the three models, the table presents the equation that characterizes them: the EoS parameter $\bar{w}(a)$ for DDE, the interaction term $Q$ for IDE, and $\Lambda(H)$ for the A1 and A2 cases of RV.
It includes explicit expressions for the dark matter density $\rho_c$, with a standard form for DDE and parameter-dependent forms for IDE and RV. 
Similarly, the dark energy density $\rho_x$ is parameter-dependent in all cases. 
The table also provides the EoS parameters, which are standard for IDE and RV, except for the DDE case where the EoS evolves dynamically. 
This structured summary facilitates a clear comparison of the mathematical frameworks and physical interpretations of the three models.

\section{Bayesian analysis}\label{sec:bayesian}

We perform a Bayesian parameter estimation to constrain the free parameters of the DDE, IDE, and RV dark energy models considered in this work. 
We use the \texttt{SimpleMC} cosmological parameter estimation code \cite{simplemc}, coupled with a nested sampling algorithm from the \texttt{dynesty} library \cite{Speagle_2020}. 
Our analysis is based on data from Type Ia supernovae (SNeIa), Baryon Acoustic Oscillation (BAO) measurements, and cosmic chronometers (CC), which are detailed below:

\begin{itemize}
    \item \textbf{SNeIa}: We use the Pantheon+ compilation \cite{Brout:2022vxf}, which consists of 1550 Type Ia supernovae covering the redshift range $z = 0.001$ to $z = 2.26$.
    \item \textbf{BAO}: We use high-precision BAO measurements at various redshifts up to $z < 2.36$, including data from BOSS DR14 quasars (eBOSS) \cite{eBOSS:2017cqx}, SDSS DR12 Galaxy Consensus \cite{BOSS:2016wmc}, Ly-$\alpha$ DR14 cross-correlation \cite{eBOSS:2019qwo}, Ly-$\alpha$ DR14 auto-correlation \cite{eBOSS:2019ytm}, the Six-Degree Field Galaxy Survey (6dFGS) \cite{Beutler_2011}, and the SDSS Main Galaxy Sample (MGS) \cite{Ross:2014qpa}.
    \item \textbf{Cosmic chronometers}: We use 31 cosmic chronometer measurements, which provide direct estimates of the Hubble parameter $H(z)$ from galaxies that evolve slowly \cite{Jimenez:2003iv, Simon:2004tf, Daniel_Stern_2010, Moresco:2012by, Zhang:2014, Moresco:2015cya, Moresco:2016mzx, Ratsimbazafy:2017vga} .
\end{itemize}

Throughout the analysis, we assume a flat FLRW universe and adopt flat priors for the free parameters: $\Omega_m \in [0.1, 0.5]$ and $h \in [0.4, 0.9]$ for the reduced Hubble constant. In addition, for the IDE model we take $\gamma \in [-0.5, 0.5]$, and for the RV model $\alpha_1 \in [-0.5, 0.5]$.

The main results of our Bayesian analysis are summarized in Table \ref{tab:parameter_estimation}.
We report the posterior parameter estimation for the matter density parameter $\Omega_m$, the reduced Hubble parameter $h$, and the dark energy model parameters $\bar{w}_0 = \bar{w}(a=1)$ for the DDE model, and the interaction parameters $\gamma$ for IDE and $\alpha_1$ for RV. 
The Bayesian evidence $\log Z$ is calculated with the used nested sampling algorithm, and we also report the difference $\Delta \log Z$ relative to $\Lambda$CDM to facilitate model comparison according to Jeffrey's scale.
For all models, we estimated the standard cosmological parameters alongside the specific dark energy parameters. 
The $\Omega_m$ value for DDE is statistically consistent with that of $\Lambda$CDM (with $\Omega_m \approx 0.31$), whereas both IDE and RV models favour slightly higher values ($\Omega_m \approx 0.34$).
Similarly, for the Hubble parameter $h$, the DDE model yields a lower value ($h \approx 0.66$) compared to $\Lambda$CDM ($h \approx 0.68$), while the IDE and RV models indicate higher values ($h \approx 0.70$). 
Regarding the dark energy parameters, the DDE model shows $\bar{w}_0 = -0.9230 \pm 0.0522$, which is indicative of a quintessence-like behaviour at the current epoch. 
For the IDE and RV models, we find that the interaction parameters are statistically similar, with $\gamma \approx 0.0442 \pm 0.0259$ and $\alpha_1 \approx 0.0488 \pm 0.0288$, respectively. 
These values represent a small deviation from the $\Lambda$CDM scenario and are consistent with the relation $\gamma = \frac{\alpha_1}{1+\alpha_1}$ derived in Section \ref{subsec: equivalence rv ide}.
Notably, $\gamma$ is slightly larger than $\alpha_1$, as expected from this relation.
In terms of Bayesian evidence, the differences $\Delta \log Z$ for DDE, IDE, and RV relative to $\Lambda$CDM are $-1.50$, $-1.15$, and $-0.97$, respectively. 
According to Jeffrey's scale, these differences imply that none of the alternative dark energy models shows a statistically significant advantage or disadvantage over $\Lambda$CDM based solely on the background data.

\begin{table}[h!]
    \centering
    \begin{tabular}{|c|c|c|c|c|c|}
        \hline
        \textbf{Model} & $\Omega_m$ & $h$ & model parameters
        & $\log Z$ & $\Delta \log Z$  \\
        \hline \hline
        \textbf{$\Lambda$CDM}  & $0.3189 \pm 0.0122$ & $0.6826 \pm 0.0081$ & -- & $-722.5305 \pm 0.1558$ & $0.0$\\
        \hline
        \textbf{DDE}  & \(0.3046 \pm 0.0153\) & \(0.6638 \pm 0.0155\) & \(\bar{w}_0 = -0.9230 \pm 0.0522\) & \(-724.0299 \pm 0.1821\) & \(-1.50\)  \\
        \hline
        \textbf{IDE}   & $0.3405 \pm 0.0177$ & $0.7049 \pm 0.0156$ & $\gamma = 0.0442 \pm 0.0259$ & $-723.6851 \pm 0.1808$ & $-1.15$\\
        \hline
        \textbf{RV} & $0.3411 \pm 0.0182$ & $0.7055 \pm 0.0157$ & $\alpha_1 = 0.0488 \pm 0.0288$ & $-723.4969 \pm 0.1780$ & $-0.97$ \\
        \hline
    \end{tabular}
    \caption{Parameter estimation for the dark energy models. Analysing the Bayes factor  ($\Delta \log Z$ ) and according to the Jeffreys' Scale, all of these models have an inconclusive advantage or disadvantage over $\Lambda$CDM.}
    \label{tab:parameter_estimation}
\end{table}

The parameter estimation indicates that while the DDE, IDE, and RV models yield slightly different values for $\Omega_m$ and $h$ compared to $\Lambda$CDM, these differences are within the error bars and the models remain statistically competitive.
The modest shifts in the dark energy parameters suggest subtle departures from the $\Lambda$CDM paradigm; however, the Bayesian evidence does not favour any one model decisively. 

\begin{figure}[t!]
    \centering
    \includegraphics[width=0.5\linewidth]{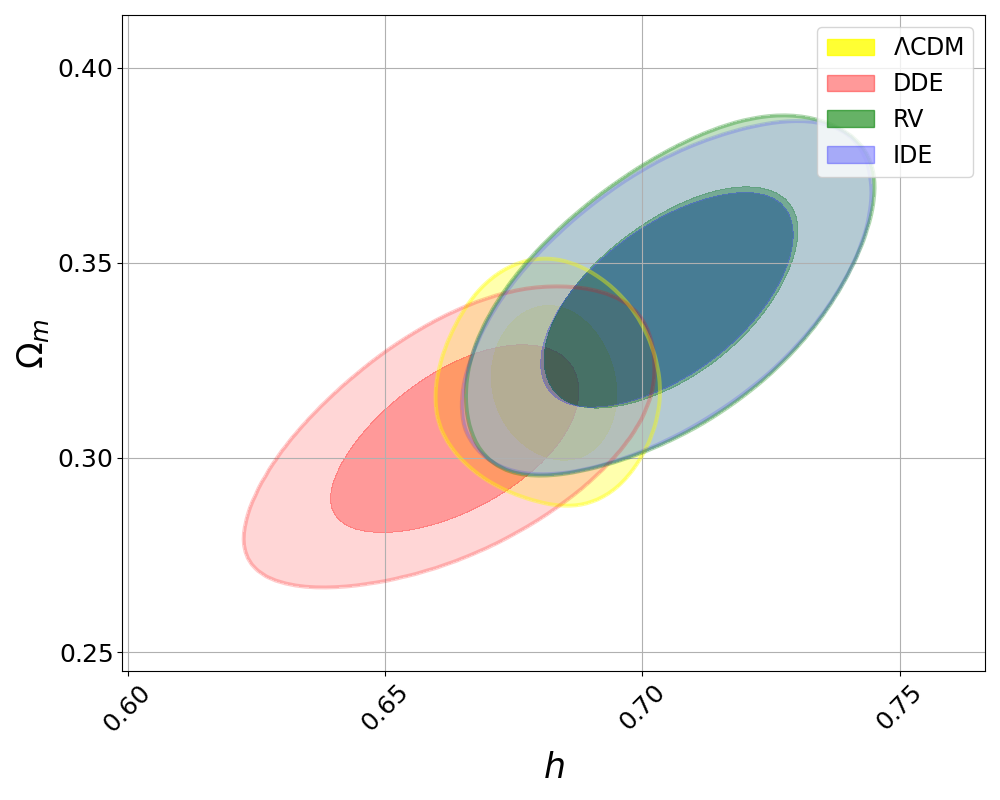}
    \caption{Two-dimensional (68\% and 95\% CLs) marginalized posterior distributions for the dark matter energy density $\Omega_m$ of $\Lambda$CDM (yellow), DDE (red), IDE (blue) and RV green.}
    \label{fig:Om h}
\end{figure}

Figure \ref{fig:Om h} shows the two-dimensional posterior distributions for the parameters $h$ and $\Omega_m$ for each dark energy model, based on the Pantheon+, BAO, and CC datasets. 
The inner and outer contours represent the 68\% and 95\% confidence levels (CLs), respectively. 
All three dark energy models display a positive correlation between $h$ and $\Omega_m$. 
Notably, the IDE and RV models favour higher values of both $\Omega_m$ and $h$ compared to $\Lambda$CDM, while the DDE model tends toward lower values. Despite these differences, the confidence contours of all models overlap with those of $\Lambda$CDM, indicating consistency with the standard cosmological model.

\begin{figure}[t!]
    \centering
    \begin{minipage}[b]{0.48\textwidth}
        \centering
        \includegraphics[width=\textwidth]{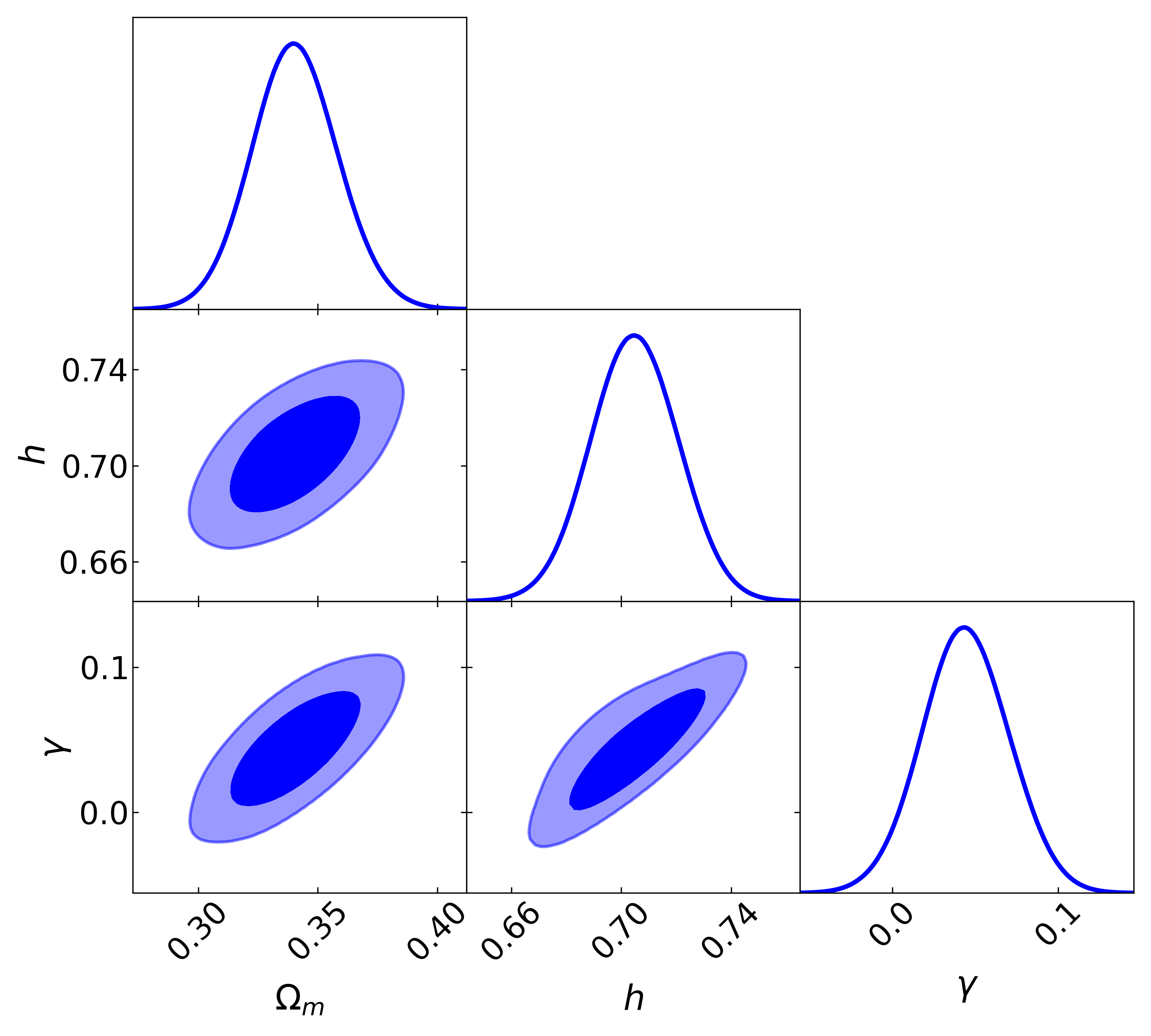}
    \end{minipage}
    \hfill
    \begin{minipage}[b]{0.48\textwidth}
        \centering
        \includegraphics[width=\textwidth]{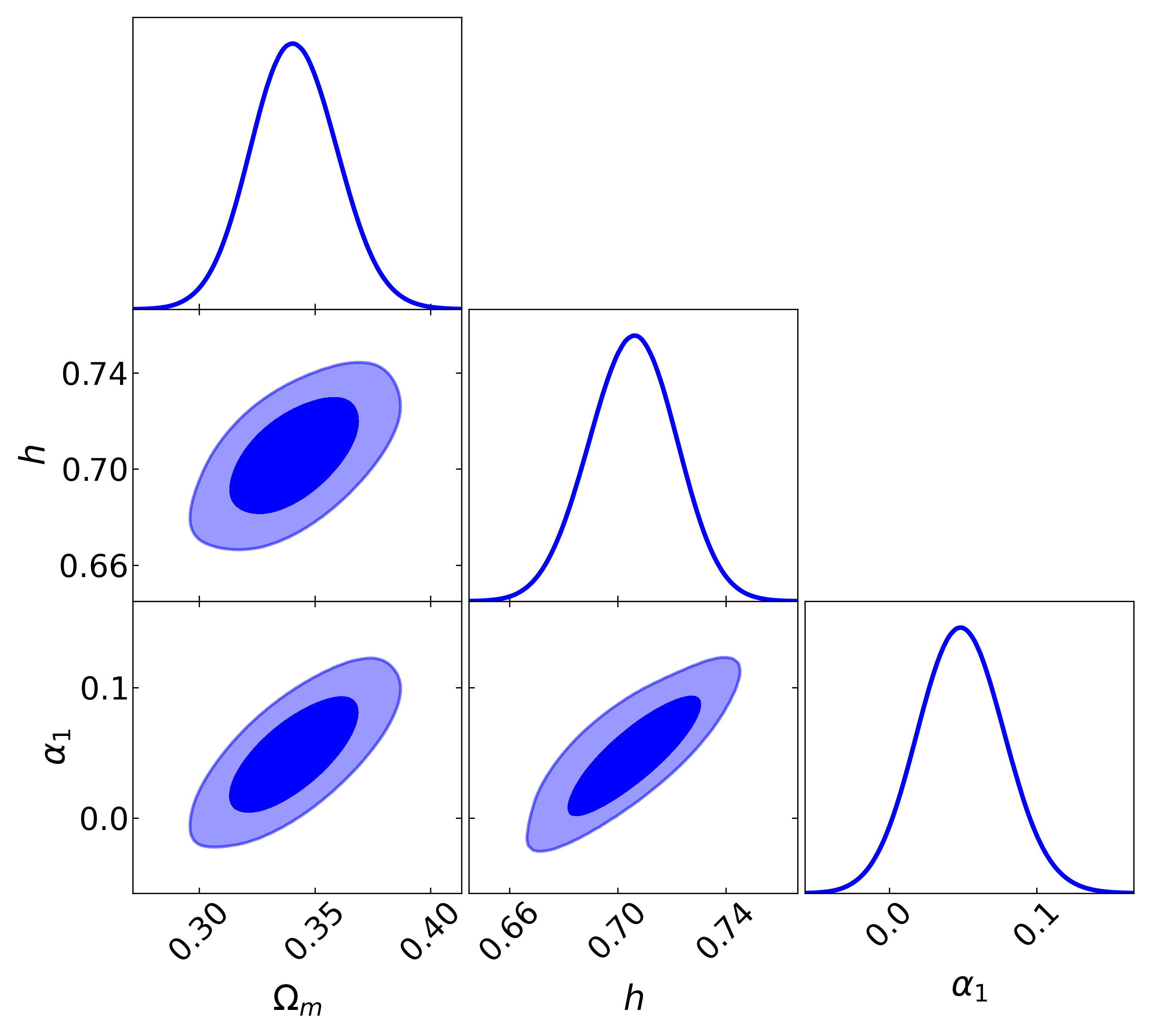}
    \end{minipage}
    \caption{Posterior plots of IDE and RV cosmological models.
    One- and two-dimensional (68\% and 95\% CLs) marginalized posterior distributions for the free parameter $\gamma$ of IDE (left) and $\alpha_1$ RV (right), using Pantheon+, BAO and CC.}
    \label{fig:corner_plots}
\end{figure}

Figure \ref{fig:corner_plots} shows the one- and two-dimensional marginalized posterior distributions for the free parameters: $\gamma$ for the IDE model (left panel) and $\alpha_1$ for the RV model (right panel), based on the Pantheon+, BAO, and CC datasets. 
Notably, both models yield nearly identical constraints, with the data favouring small positive values on the order of $10^{-2}$. 
Given the relation $\gamma = \frac{\alpha_1}{1+\alpha_1}$, this result implies that $\gamma \gtrsim \alpha_1$, in agreement with our earlier theoretical discussion. 
Additionally, the parameters $\gamma$ and $\alpha_1$ exhibit a positive correlation with both $\Omega_m$ and $h$, underscoring their interdependence within the dark energy models.

As a reference, constraints for the IDE and RV models obtained in previous studies are summarized in Table \ref{tab:parameters_constraints}. 
For the observational constraints on the IDE model, described by equation \eqref{IDE Q}, $\tilde{Q} = 3H\gamma \tilde{\rho}_c$, we reference the results from \cite{Costa:2016tpb}, where a global fit to Planck, BAO, SNIa from Joint Light Analysis (JLA) \cite{SDSS:2014iwm} and $H0$ data yielded $\gamma = 0.0007127^{+0.000256}_{-0.000633}$ at 68\% CL. 
Similarly, \cite{Hoerning:2023hks} analysed using CMB, BAOS, SNIa, and RSD data, obtaining $\gamma = 0.00088^{+0.00088}_{-0.00068}$ at 68\% CL. 
These results indicate that $\gamma$ is constrained to the range $\sim 10^{-3}-10^{-4}$. 
For the observational constraints on the RV A2 model, described by equation \eqref{A2}, $\Lambda(H) = a_0 + a_1 \dot{H} + a_2 H^2$, we reference the work of \cite{SolaPeracaula:2021gxi}. 
Using SNIa, BAO, CC, LSS, and CMB data, they found the parameter value in their notation to be $\nu_{eff} = 0.00024^{+0.00039}_{-0.00040}$ at 68\% CL. 
Here, $\nu_{eff} \simeq \frac{\nu}{4}$, which, when converted to the notation used in this work, corresponds to $\nu = \frac{2}{3} a_1 = \frac{1}{3} a_2$. 
From this, we obtain $\alpha_1 = 0.00024$. 
Similarly, \cite{SolaPeracaula:2023swx}, using the same dataset combination, found $\nu_{eff} = 0.00006 \pm 0.00030$, which translates to $\alpha_1 = 0.00006$. 
Therefore, the parameter $\alpha_1$ can be constrained to the range $\sim 10^{-4}-10^{-5}$.

\begin{table}[h!]
    \centering
    \begin{tabular}{|c|c|c|}
        \hline
        \textbf{Model} & parameter constraint & reference \\
        \hline \hline
        IDE $Q = 3H\gamma \tilde{\rho}_c$ & $\gamma \sim 10^{-3}-10^{-4}$ & \cite{Costa:2016tpb, Hoerning:2023hks} \\ 
        \hline
        RV Model A2 & $\alpha_1 \sim 10^{-4}-10^{-5}$ & \cite{SolaPeracaula:2021gxi, SolaPeracaula:2023swx} \\
        \hline
    \end{tabular}
    \caption{Parameter constraints for the IDE and RV models. The table lists the parameter ranges for the interaction term $\gamma$ in the IDE model and the coefficient $\alpha_1$ in the RV A2 model, along with their corresponding references.}
    \label{tab:parameters_constraints}
\end{table}

Although previous studies have incorporated additional datasets, including CMB and LSS measurements, our analysis focuses on background observables and relies on the updated Pantheon+ compilation, BAO, and CC.
These datasets provide robust constraints on the expansion history, which are sufficient to confirm the theoretical equivalence among DDE, IDE, and RV models.
Moreover, by using the most recent Pantheon+ dataset, an improvement over the earlier Pantheon compilation \cite{Pan-STARRS1:2017jku}, we ensure that our constraints reflect the latest supernova observations, further validating the robustness of our findings. 
This consistency highlights that the established model equivalences are not contingent on specific dataset choices but rather represent fundamental relationships in the description of the dark sector.

\section{Conclusions}\label{sec: conclusions}

In this paper, we have explored the degeneracy in the dark sector, demonstrating the equivalence between dynamical dark energy (DDE), interacting dark energy (IDE), and running vacuum (RV) models at the background level. 
We began by summarizing the general mapping between DDE, characterized by the dark energy EoS parameter $\bar{w}(a)$, and IDE, described by a scalar interaction term $Q$, as previously shown in \cite{vonMarttens:2019ixw}. 
We then extended the study by first showing that the A1 and A2 RV models, characterized by $\Lambda(H) = a_0 + a_1 \dot{H} + a_2 H^2$, are equivalent to IDE models with an interaction term of the form $Q = 3H\gamma \hat{\rho}_c$. 
Subsequently, using the DDE-IDE relations, we constructed the specific form of $\bar{w}(a)$ for a DDE model equivalent to the studied RV models. 
These steps explicitly demonstrate that one model can be mathematically transformed into another, providing a unified framework for describing the dark sector.
To validate these theoretical equivalences, we conducted a Bayesian parameter estimation analysis for all three models using background data from SNIa, BAO and CC. 
It is important to note that the Bayesian analysis was not intended to compare the performance of these models against $\Lambda$CDM, as there are already several detailed studies addressing parameter constraints for DDE, IDE, and RV. 
Instead, the primary goal was to complement the theoretical findings by demonstrating consistency with observational constraints on the dark energy parameters involved.
In summary, the equivalence between the three dark energy models derived theoretically is consistent with observational data, reinforcing the validity of this unified perspective on the dark sector. 
This contributes to the convergence and unification of the description of the dark sector.

The DDE-IDE equivalence is general and encompasses a wide range of models. 
This is because the transformation equations rely solely on the energy densities of the dark sector in both models and the EoS of the DDE, without requiring additional assumptions about the cosmological framework. 
In contrast, the IDE-RV equivalence established in this work relies on two key assumptions. 
First, we restrict the analysis to the A1 and A2 RV models, effectively considering only the leading terms in the general expression for $\Lambda(H)$ shown in equation \eqref{rv general}, $\Lambda = a_{0} + \sum_{k} b_{k} H^{2k} + \sum_{k} c_{k} \dot{H}^{k} \approx a_0 + a_1 H^2 + a_2 \dot{H}$. 
Second, we assume that the dark sector comprises the entirety of the cosmic budget in the RV cosmological model, such that $\hat{\rho}_{tot} = \hat{\rho}_c + \hat{\rho}_x$. 
These assumptions are crucial for deriving equation \eqref{rv de}, $\hat{\rho}_x = \alpha_0 + \alpha_1 \hat{\rho}_c$. 
If, for example, radiation were included, equation \eqref{rv de} would no longer hold, as the radiation energy density would contribute additional terms. 
It is important to note that the IDE-RV equivalence presented in this work is not necessarily unique. 
More general (albeit more complex) IDE-RV equivalences may exist. 
The same argument applies to the RV-DDE equivalence. 
Nonetheless, the $\Lambda = a_0 + a_1 H^2 + a_2 \dot{H}$ RV model has been shown to be a viable dark energy candidate. 
When considering a cosmology dominated by the dark sector, it provides an adequate description of the late-time universe, making it a suitable framework.
A natural next step following this work is to extend the investigation of equivalences between DDE, IDE, and RV models to the perturbation level, focusing on how the degeneracy observed at the background level is broken.
Additionally, future work could aim to generalize the IDE-RV equivalence presented in this study, developing a more comprehensive framework that relaxes the assumptions made here. 
Expanding the scope to include other dark energy models, such as the generalized Chaplygin gas or holographic dark energy, could further broaden the range of equivalences and uncover new connections between these scenarios. 

As a final comment, the background equivalence of dark energy models implies that two or more models, inspired by different physical motivations, can be indistinguishable when constrained by observational data that depends solely on the evolution of the Hubble rate, such as SN or BAO measurements. 
Among the plethora of proposed dark energy models, it is possible that many of them are equivalent or nearly equivalent at the background level, describing the same Hubble expansion evolution while the dark sector evolves in different ways. 
This highlights the importance of recognizing and exploring the extent to which different models may, in fact, describe the same underlying physics, even when presented as novel proposals.
While such analyses may be more complex and demanding, they provide valuable insights by emphasizing the role of perturbative studies in breaking the degeneracy and advancing our understanding of the dark sector.
Therefore, pursuing convergence in the description of dark energy models is a crucial complementary approach to unravelling the mystery of the dark sector, while also giving due importance to perturbative studies for refining our understanding of these models.
A systematic effort to identify and streamline equivalent or redundant models is essential to focus research efforts and pave the way for meaningful progress in this field.

\section{Acknowledgments}

We are grateful to Dr. J. Alberto Vázquez for helpful discussions. IGV thanks the University of Geneva. 

\bibliography{references}
\end{document}